\documentclass[twoside,journal]{IEEEtran}
\usepackage{amsmath,amssymb,amsfonts}
\usepackage{cite}
\usepackage{algorithmic}
\usepackage{graphicx}
\usepackage{multirow}
\usepackage{textcomp}
\usepackage{stfloats}
\usepackage{url}
\usepackage{verbatim}
\usepackage{balance}

\makeatletter
\newcommand{\thickhline}{%
    \noalign {\ifnum 0=`}\fi \hrule height 1pt
    \futurelet \reserved@a \@xhline
}
\usepackage[colorlinks=true, allcolors=black,urlcolor = black]{hyperref}
\DeclareFontFamily{U}{wncy}{}
\DeclareFontShape{U}{wncy}{m}{n}{<->wncyr10}{}
\DeclareSymbolFont{mcy}{U}{wncy}{m}{n}
\DeclareMathSymbol{\Sh}{\mathord}{mcy}{"58} 

\def\BibTeX{{\rm B\kern-.05em{\sc i\kern-.025em b}\kern-.08em
    T\kern-.1667em\lower.7ex\hbox{E}\kern-.125emX}}
\begin{document}
\title{Uni-COAL: A Unified Framework for Cross-Modality Synthesis and Super-Resolution of MR Images}
\author{Zhiyun Song, Zengxin Qi, Xin Wang, Xiangyu Zhao, Zhenrong Shen, Sheng Wang, Manman Fei, Zhe Wang, Di Zang, Dongdong Chen,  Linlin Yao, Qian Wang, Xuehai Wu, Lichi Zhang
\thanks{Z. Song and Z. Qi contributed equally to this manuscript.}
\thanks{Corresponding authors: L. Zhang (e-mail: lichizhang@sjtu.edu.cn) and Xuehai Wu (e-mail: wuxuehai2013@163.com).}
\thanks{Z. Song, X. Wang, X. Zhao, Z. Shen, S. Wang, M. Fei, D. Chen, L. Yao, and L. Zhang are with the School of Biomedical Engineering, Shanghai Jiao Tong University, Shanghai, 200030, China (e-mails: zhiyunsung@gmail.com, \{wangxin1007, xiangyu.zhao, zhenrongshen, wsheng, feimanman, chendongdong, yaolinlin23, lichizhang\}@sjtu.edu.cn).}
\thanks{Z. Qi, Z. Wang, D. Zang, and X. Wu are with Department of Neurosurgery, Huashan Hospital, Shanghai Medical College, Fudan University, National Center for Neurological Disorders, Shanghai Key Laboratory of Brain Function and Restoration and Neural Regeneration, State Key Laboratory of Medical Neurobiology and MOE Frontiers Center for Brain Science, School of Basic Medical Sciences and Institutes of Brain Science, Fudan University (e-mails: qizengxin@huashan.org.cn, \{20211220121, dzang16\}@fudan.edu.cn, wuxuehai2013@163.com).}
\thanks{Q. Wang is with School of Biomedical Engineering, ShanghaiTech University, Shanghai 201210, China, 
and with Shanghai Clinical Research and Trial Center, Shanghai 201210, China (e-mail: qianwang@shanghaitech.edu.cn).}
}

\maketitle

\begin{abstract}
Cross-modality synthesis (CMS), super-resolution (SR), and their combination (CMSR) have been extensively studied for magnetic resonance imaging (MRI).
Their primary goals are to enhance the imaging quality by synthesizing the desired modality and reducing the slice thickness. 
Despite the promising synthetic results, these techniques are often tailored to specific tasks, thereby limiting their adaptability to complex clinical scenarios.
Therefore, it is crucial to build a unified network that can handle various image synthesis tasks with arbitrary requirements of modality and resolution settings, 
so that the resources for training and deploying the models can be greatly reduced. 
However, none of the previous works is capable of performing CMS, SR, and CMSR using a unified network.
Moreover, these MRI reconstruction methods often treat alias frequencies improperly, resulting in suboptimal detail restoration.
In this paper, we propose a Unified Co-Modulated Alias-free framework (\textit{Uni-COAL}) to accomplish the aforementioned tasks with a single network.
The co-modulation design of the image-conditioned and stochastic attribute representations ensures the consistency between CMS and SR, 
while simultaneously accommodating arbitrary combinations of input/output modalities and thickness.
The generator of \textit{Uni-COAL} is also designed to be alias-free based on the Shannon-Nyquist signal processing framework, ensuring effective suppression of alias frequencies.
Additionally, we leverage the semantic prior of Segment Anything Model (SAM) to guide \textit{Uni-COAL}, ensuring a more authentic preservation of anatomical structures during synthesis.
Experiments on three datasets demonstrate that \textit{Uni-COAL} outperforms the alternatives in CMS, SR, and CMSR tasks for MR images, which highlights its generalizability to wide-range applications.

\end{abstract}


\section{Introduction}

\begin{figure}[t]
\includegraphics[width=\columnwidth]{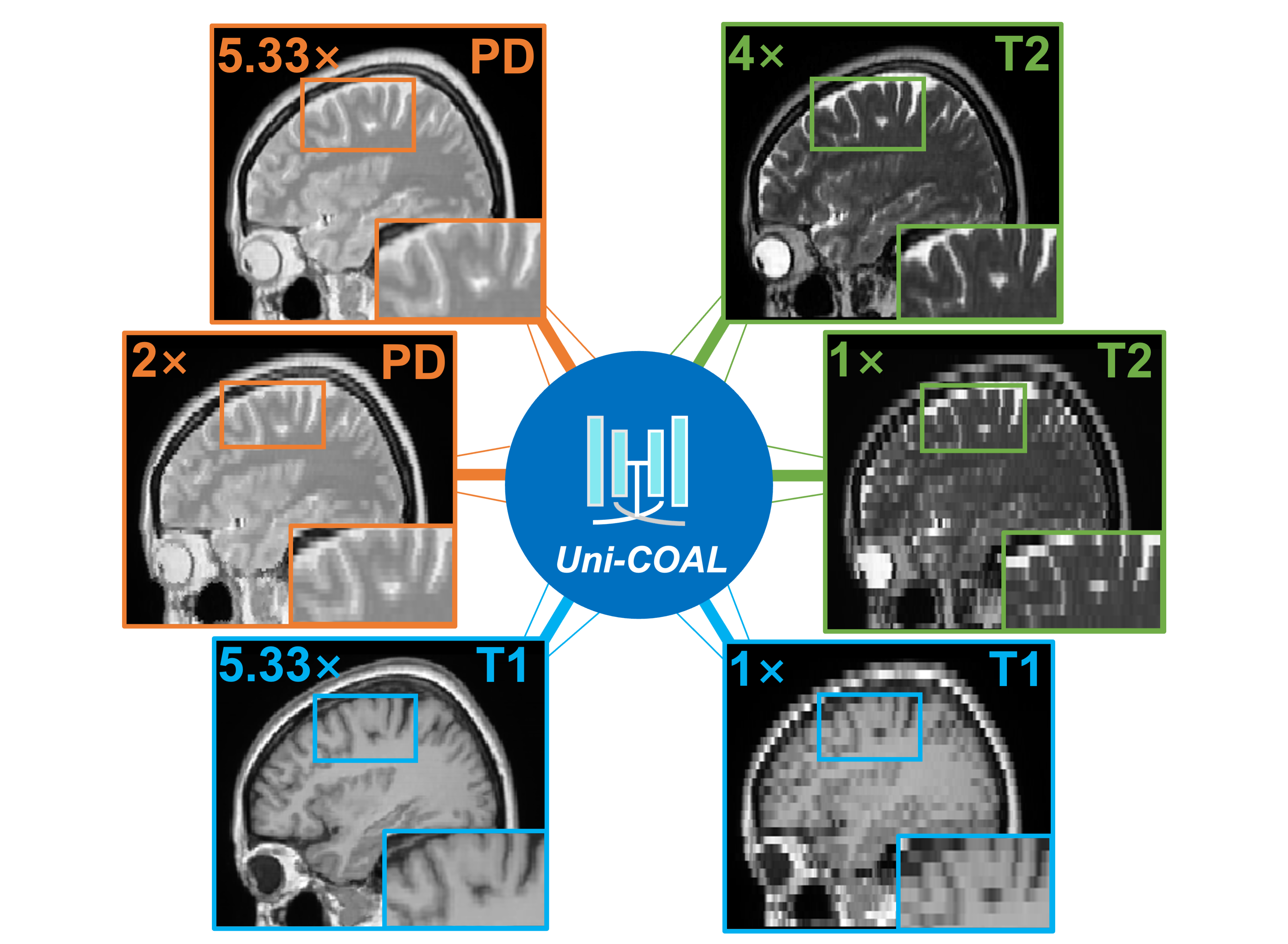}
\caption{
Qualitative results of \textit{Uni-COAL} for the arbitrary cross-modality synthesis and super-resolution of MR images on the sagittal view. \textit{Uni-COAL} is capable of synthesizing target MR images with varying modalities and slice thickness using a \textbf{single} network.
Note that the MR images with spatial resolution of 
0.9375 mm $\times $ 0.9375 mm $\times $ 5.0 mm are labeled as "1$\times$". 
} \label{intro}
\end{figure}

\IEEEPARstart{M}{agnetic} Resonance Imaging (MRI) is widely recognized as an important neuroimaging technique that provides rich information about brain tissue anatomy in a non-invasive manner. Several studies have demonstrated that multi-modal MR images offer complementary information on tissue morphology, enabling a more comprehensive brain analysis \cite{brats,ResViT}. For example, T1-weighted images are used to distinguish between gray and white matter tissues, while FLAIR images provide clear observations of edematous regions from cerebrospinal fluid \cite{contrast_func}. Moreover, 3D MR imaging with isotropic voxels yields more detailed information than its 2D counterparts with anisotropic resolution, thereby facilitating more accurate analysis of the anatomical structure\cite{sr_significance1}.

Although the multi-modal and isotropic acquisition of MR images has been extensively applied in research studies, clinical scanning protocols are generally much inferior due to the high cost of scanning resources and long acquisition time\cite{ResViT, DeepResolve, synthsr}.
In clinical settings, physicians typically adopt two strategies to simplify the scanning procedures:
The first one involves scanning a specific modality that can reveal the patient's brain situation at minimal costs, rather than acquiring the full modalities.
The second strategy involves the adoption of 2D scanning protocols that scan anisotropic images with high in-plane resolution but large slice thickness (\emph{e.g.} MR image in Fig.\ref{intro} with in-plane resolution of 0.9375 mm $\times $ 0.9375 mm and thickness of 5.0 mm).

However, these clinical images acquired with potentially a single modality and high slice thickness are unsuitable for further research studies due to the diminished quality.
Therefore, image enhancement techniques have been developed with the aim of enhancing the quality of these MR images. This is achieved through the synthesis of alternative modalities and the implementation of super-resolution techniques. The corresponding techniques can be categorized as cross-modality synthesis (CMS), super-resolution (SR), and simultaneous cross-modality synthesis and super-resolution (CMSR).
Most CMS works focus on modeling modality-invariant details to facilitate translation between modalities, and are typically implemented for 2D images \cite{CMS_CNN0, CMS_CNN2, ResViT, CMS_Diff0}.
Later, 3D-based \cite{CMS_GAN_3D0, SCGAN} or 2.5D-based \cite{pGAN, ProvoGAN} CMS methods have also been proposed to ensure coherence among the generated slices by incorporating neighboring slices as context information.
In contrast to CMS methods, SR methods \cite{DCSRN, DeepResolve, SR_GAN0} typically employ 3D models, given the crucial role of slice correlation and the requirement for consistent quality across different views. Furthermore, CMSR methods aim to implement both the cross-modality and super-resolution tasks for the target images. 
However, studies \cite{WEENIE} show that simply performing CMS and SR sequentially in pursuit of CMSR demonstrates inferior results, which may stem from error accumulation.
Therefore, some attempts \cite{synthsr, WEENIE} propose to perform CMSR using a joint model. 


Despite their success in improving the quality of acquired MR images, these works have two main limitations that hinder their applications in clinical practice:
First, they are generally tailored to specific tasks, which restricts their applications in an arbitrary manner. 
As MR imaging in clinical scenarios often involves various combinations of modalities and resolutions, building specialized models for every specific setting is time-consuming and resource-intensive. Therefore, it is more desirable to use a single and universal network to handle these tasks flexibly.
For example, AutoGAN \cite{AutoGAN} was proposed to automatically search for optimal generator architectures for CMS with different input/output modality combinations.
Another notable approach that addresses this requirement is the arbitrary scale super-resolution (ASSR) \cite{ArSSR, ASSR1, ASSR3}, which is designed to reduce the thickness of MR images to arbitrary values.
However, these approaches are confined to the specialized image synthesis tasks, \emph{i.e.}, CMS or SR, which still hinders their applications in the clinical scenario.
To the best of our knowledge, none of the previous works can simultaneously handle various combinations of modalities and resolutions with a unified model.

\begin{figure}[t]
\includegraphics[width=\columnwidth]{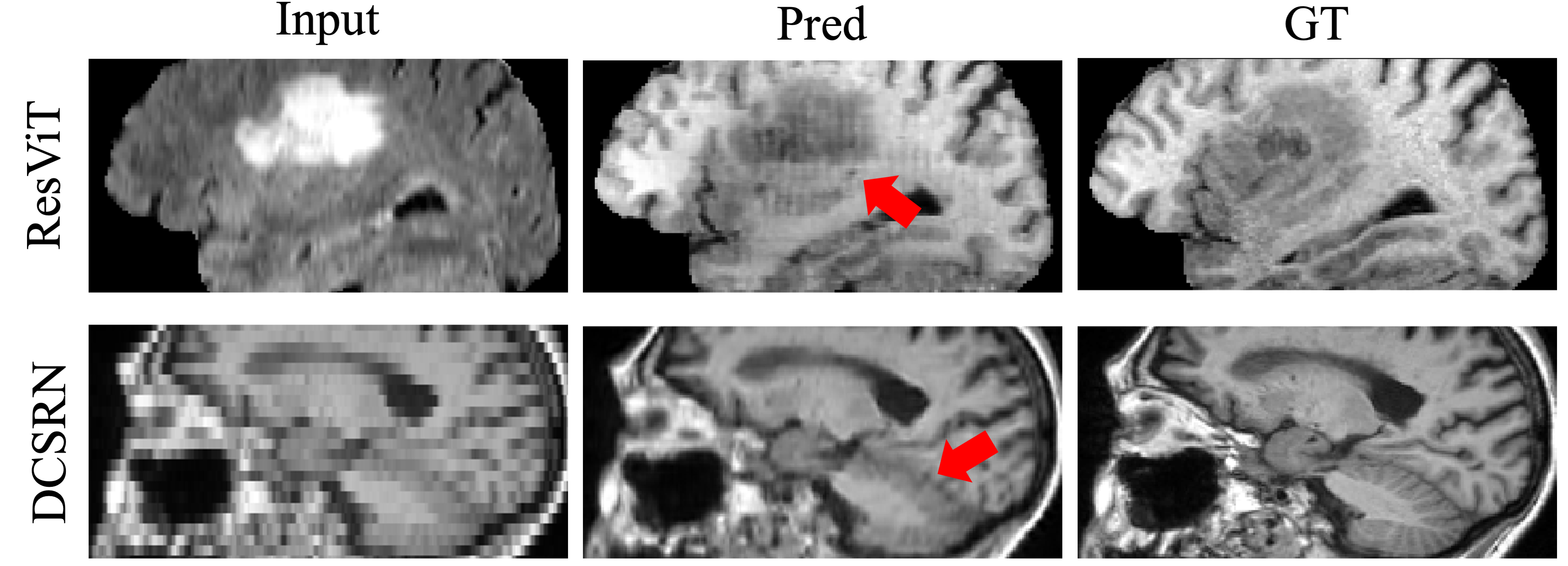}
\caption{
The synthetic results produced by the previous CMS (ResViT) and SR (DCSRN) methods. All of the generated MR images suffer from artifacts, as compared with the reference ground truth.
} \label{aliasing}
\end{figure}

Another limitation of these deep learning-based methods is the inadequate treatment of alias frequencies.
Aliasing is a prevalent phenomenon in the MRI reconstruction field, especially in 2D MRI acquisitions with low resolution in the through-plane direction, where aliasing artifacts may arise due to Nyquist criteria \cite{SMORE}.
Additionally, aliasing effects should be carefully considered when reconstructing from undersampled k-space data \cite{aliasingMRI}.
However, current works for MRI synthesis often overlook the impact of aliasing, despite treating the signals within the neural network as discrete entities.
Consequently, these works produce unnatural details, and even aliasing artifacts as shown in Fig. \ref{aliasing}.

In this paper, we propose a Unified Co-Modulated Alias-free framework (\textit{Uni-COAL}) to address the challenges above. 
To achieve flexible CMS, SR, and CMSR with a unified model, 
we propose a co-modulation design, which aims to extend style-based modulation \cite{StyleGAN2} with image-conditioned and stochastic attribute-embedded representations. Such design ensures consistency between different tasks and allows us to generate slices of any desired modality at any position, enabling arbitrary-modality CMS and arbitrary-scale SR, as depicted in Fig. \ref{intro}.
Moreover, we alleviate the problem of aliasing artifacts by considering the feature maps in the neural network as continuous signals.
By designing the discrete operators in the generator to be alias-free under the Shannon-Nyquist signal processing framework, \textit{Uni-COAL} effectively suppresses the aliasing artifacts and restores high-frequency details more naturally.
Additionally, we propose to guide the generator with the Segment Anything Model (SAM) \cite{SAM}, a powerful foundational model for general image segmentation. 
Here we utilize the SAM model by extracting the semantic features of the target image to represent its anatomical structure, which provides additional prior information for \textit{Uni-COAL}, and helps it generate precise anatomical structures.

The contributions of this paper are summarized as follows:
\begin{enumerate}
    \item We propose a unified framework named \textit{Uni-COAL} to accomplish CMS, SR, and CMSR uniformly, which facilitates MRI synthesis with various combinations of modality and slice thickness settings.
    \item The generator of \textit{Uni-COAL} is designed to be alias-free, which helps restore high-frequency details more naturally for the reconstructed images. 
    \item We incorporate the SAM model when training the \textit{Uni-COAL}, which can provide prior knowledge about the anatomical structure of the generated MR images.
    \item Comprehensive experiments are conducted on several datasets, which validates that \textit{Uni-COAL} achieves competitive performances on CMS, SR, and CMSR tasks.
\end{enumerate}

\section{Related Work}
\subsection{Cross-modality Synthesis}
Cross-modality synthesis adopts image-to-image translation techniques to reconstruct the desired modality \cite{CMS_overview}.
Early studies in this field used sparse-coding and example-based methods \cite{CMS_sparse0, CMS_sparse1}, where each patch with source modality is matched to a combination of relevant patches in the learned dictionary.
Later, CNN-based methods\cite{CMS_CNN0, CMS_CNN2} overcome the problems brought by patch-based methods and yield more robust results than traditional ones. 
To further improve the synthesis quality with refined structural details, generative adversarial networks (GAN) and diffusion model (DM) are introduced to learn the distribution of the target modality comprehensively. Given the paired images with the original and target modalities, these methods utilize pix2pix-based framework to generate target MR images conditioned on the input images.
Later on, these CMS approaches are further improved with modified loss functions \cite{pGAN, CMS_Diff0}, enhanced architectures \cite{ResViT, CMS_arch}, or 3D-based generation \cite{CMS_GAN_3D0, SCGAN}.
Among them, 3D-based CMS utilizes multi-dimensional spatial information, which helps produce target modalities with higher consistency than 2D-based methods.
However, they require higher computational resources to deal with 3D images and are only eligible to handle near-isotropic images, which limits their further applications in clinical practice where large proportions of acquired MR images are anisotropic.
In this work, we propose to empower the computing-efficient 2D-based CMS with spatial information by incorporating neighboring planes as context information. Although this cross-sectional strategy has been applied in previous work \cite{pGAN} for near-isotropic MRI synthesis, our framework deals with more flexible situations where either isotropic or anisotropic volumes can be taken as inputs.

\subsection{Super-resolution}
In many clinical scenarios, the acquired MRI exhibits high in-plane resolution but large thickness, resulting in a loss of inter-slice details and potential misinterpretations of the acquired images. Therefore, super-resolution techniques have been developed with the aim of reducing the slice thickness to a desired level.
Traditional optimization-based methods design a priori-guided image degradation model, which reconstructs high-resolution (HR) images using fidelity-based and regularization-based objective functions \cite{SR_tradition0, SR_tradition1}.
Recently, deep learning has become the mainstream for SR, with 3D-based CNN \cite{DCSRN, DeepResolve} achieving state-of-the-art results.
Similar to CMS, GAN is also introduced to enhance grain texture and add structure realism for SR \cite{SR_GAN0}.
Apart from these approaches that solely perform fixed-scale SR, another important category of SR is arbitrary-scale super-resolution (ASSR), which aims to accommodate thickness reduction requirements in various scale settings using a single model. 
For instance, SAINT \cite{ASSR1} achieves ASSR in both the sagittal and coronal views with a modified 2D Meta-SR model \cite{metaSSR}, followed by the refinement process with a 3D fusion network.
ArSSR \cite{ArSSR} addresses the subject-specific limitation of 3D MRI ASSR by incorporating a novel implicit voxel function.
However, these methods suffer from blurriness and artifacts, which can be attributed to the incorrect handling of details.
Some attempts aim to utilize the DM technique for synthesizing high-resolution MR images with refined details \cite{ASSR3}. However, the generation time of current DM-based synthesis methods is often reported to be excessively long, making them impractical for clinical deployment.
In this paper, we propose an alias-free network that restores fine-grained details efficiently.
Note that our method can be easily extended for ASSR with the co-modulation design.

\subsection{Simultaneous Cross-modality Synthesis and Super-resolution}
Considering the scenario where both CMS and SR need to be performed, a straight solution is to sequentially perform CMS and SR to get the results.
However, experiments in \cite{WEENIE} show that it fails to produce satisfactory results, potentially due to the accumulation of error.
Simultaneous cross-modality synthesis and super-resolution (CMSR) can tackle this problem by using a single model to replace the two sequential models.
Early work attempts to implement CMSR using sparse dictionary learning operated on the whole image regions \cite{WEENIE}.
Recent studies have utilized convolutional neural networks (CNNs) to achieve the desired CMSR task. 
For example, 
SynthSR \cite{synthsr} designed a 3D UNet-based model to reconstruct MP-RAGE volumes with thickness of 1 mm from input with various modalities and resolutions.
They also proposed to use synthetic images with perturbed resolutions and intensities to accomplish the task in an unsupervised manner, while exhibiting slightly inferior performance compared to the supervised method.
However, these works solely produce HR images with static modality and thickness, limiting their wider applications.
This study presents a highly flexible framework that performs CMSR with arbitrary target modality and slice thickness, which can be adapted to various clinical applications.


\section{Methods}
\label{sec:methods}

In this section, we commence by presenting the problem formulation of general MRI image enhancement in Sec. \ref{sec:problem_form}.
In Sec. \ref{sec:overview}, we present the overall pipeline of \textit{Uni-COAL}.
Then, we introduce the major components in the framework, including the co-modulation design in Sec. \ref{sec:co-modulation}, the alias-free generator in Sec. \ref{sec:alias-free}, the projection discriminator in Sec. \ref{sec:projection}, and the SAM-guided strategy in Sec. \ref{sec:sam}.
Finally, we outline the training details in Sec. \ref{sec:implement}.


\begin{figure*}[t]
\includegraphics[width=\textwidth]{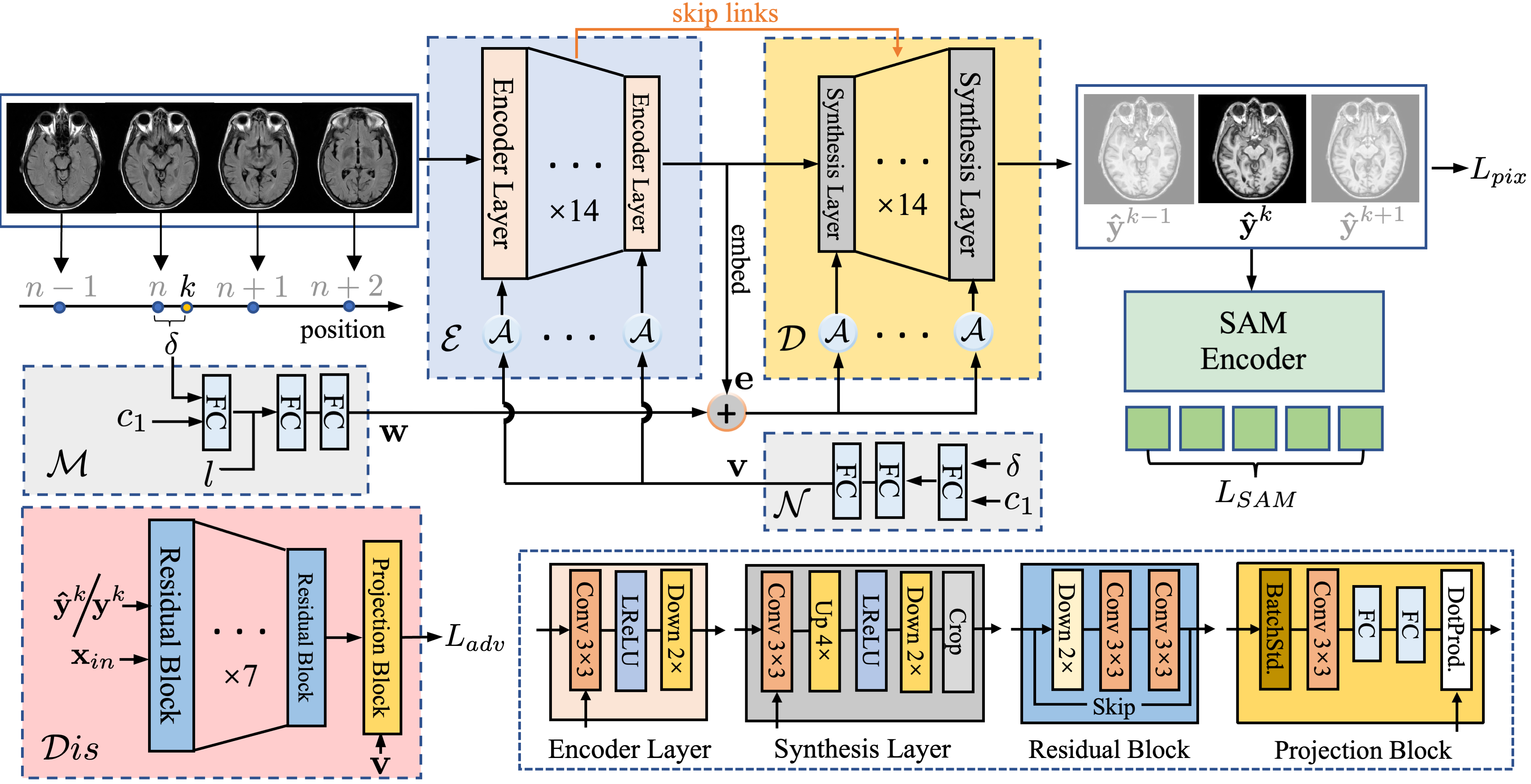}
\caption{Overall architecture of \textit{Uni-COAL}. 
The encoder $\mathcal{E}$ embeds multiple slices into image-conditioned representation \textbf{e}, while the mapping network $\mathcal{M}$ maps the latent code $l$, relative position $\delta$, and target modality $c_1$ into stochastic attribute representation $\mathbf{w}$.
The concatenated representations are transformed with affine transformation $\mathcal{A}$ and then modulate the weights of the convolution kernels of the decoder $\mathcal{D}$.
The target MR volume $\hat{\mathbf{Y}}$ is generated by traversing all target positions and concatenating the generated slices $\hat{\mathbf{y}}^k$.
The resampling filter and nonlinearity in $\mathcal{E}$ and $\mathcal{D}$ are designed to suppress the corresponding alias frequencies and ensure alias-free generation. The representation produced by the SAM encoder is used to constrain the semantics of the generated images.
} \label{overall}
\end{figure*}

\subsection{Problem Formulation}
\label{sec:problem_form}
We denote the scanned MR volume with modality $c_0$ and thickness $h_0$ as $\mathbf{X}$.
In practice, the clinical image usually has high in-plane resolution but low through-plane resolution, resulting in large $h_0$. We use $\mathbf{x}^k$ to represent the $k$-th slice along the orthogonal direction of the plane. 
Cross-modality synthesis (CMS), super-resolution (SR), or their combinations (CMSR) aim to reconstruct target MR images $\mathbf{Y}$ with modality $c_1$ and thickness $h_1$ given $\mathbf{X}$.
They accomplish CMS when $c_0$$\neq$$c_1$ and $h_0$=$h_1$, or perform SR when $c_0$=$c_1$ and $h_0$$\neq$$h_1$.
The more challenging task CMSR is performed when $c_0$$\neq$$c_1$ and $h_0$$\neq$$h_1$.
Note that for SR or CMSR, the scaling ratio $h_0/h_1$ could be any integer or non-integer value due to the arbitrary combinations of $h_0$ and $h_1$.

\subsection{Overview}
\label{sec:overview}
The overall framework of \textit{Uni-COAL} for reconstructing MR images with various combinations of modalities and resolutions is illustrated in \autoref{overall}, which is developed as GAN-based method that includes the generator and discrminator as two major parts. 
The generator consists of three main components including: 1) an image encoder $\mathcal{E}$ that learns the image-conditioned representation given original MR images, 2) a mapping network $\mathcal{M}$ that obtains the stochastic attribute representation, and 3) an image synthesizer $\mathcal{D}$ that generates the target MR images given representations from $\mathcal{E}$ and $\mathcal{M}$, which is named co-modulation in our framework.
Rather than directly generating the whole 3D volumes which can be computationally demanding and result in fixed thickness, we propose to generate 2D slices sequentially by traversing each target position, and obtain the final results by concatenating the generated slices. 
The operators in the generator are designed to be alias-free under the Shannon-Nyquist signal processing framework to ensure that details can be restored more naturally.
We also incorporate the projection discriminator design in \textit{Uni-COAL}, so that the discriminator can judge the coherence and fidelity of the generated slice.
Finally, we incorporate the SAM model which can provide the anatomical structure prior to the generator and help refine the details of ROIs in the reconstructed images.


\subsection{Co-Modulation Design}
\label{sec:co-modulation}
To generate MR images with arbitrary modality and thickness, we propose a co-modulation design to control the generation process of our synthesizer.
Our modulation information is composed of two disentangled representations: image-conditioned representation $\mathbf{e}$ and stochastic attribute representation $\mathbf{w}$.
The former one is obtained by encoding several adjacent slices $\mathbf{x}_{in}$ with the encoder $\mathcal{E}$:
\begin{equation}
\mathbf{e}=\mathcal{E}(\mathbf{x}_{in})=\mathcal{E}([\mathbf{x}^{n-(m/2-1)}, ..., \mathbf{x}^{n}, ..., \mathbf{x}^{n+m/2}]),
\label{x_in}\end{equation}
where $n=\lfloor kh_1/h_0\rfloor$ is determined by the target slice $\mathbf{y}^k$, $m$ is the number of input slices, which is set as 4 following \cite{FLAVR}. 

In the mapping network $\mathcal{M}$, the stochastic attribute representation $\mathbf{w}$ is transformed from the latent code $l\sim N(0,\mathbf{I} )$, the target modality class $c_1$, and the relative slice index $\delta=kh_1/h_0 - \lfloor kh_1/h_0\rfloor$.
Specifically, the relative slice index $\delta$ and target modality $c_1$ are embedded independently, which are then concatenated with the latent code $l$.
Next, two fully connected layers are used to transform the concatenated embeddings into the stochastic attribute representation $\mathbf{w}$ following the class-conditional design \cite{StyleGAN2-ada}.

For the image synthesizer $\mathcal{D}$, $\mathbf{e}$ and $\mathbf{w}$ are concatenated to produce a style vector $s$ with the affine transform $\mathcal{A}$:
\begin{equation}
s = \mathcal{A}(\mathbf{e}, \mathbf{w}).
\label{comodgan}\end{equation}

The style vector $s$ is then used to modulate the weights of the convolution kernels in the image synthesizer $\mathcal{D}$ via weight demodulation \cite{StyleGAN2}:
\begin{equation}
w_{i j k}^{\prime}=(s_i \cdot w_{i j k}) / \sqrt{\sum_{i, k}(s_i \cdot w_{i j k})^2+\epsilon},
\label{modulatedconv}\end{equation}
where $w_{i j k}$ and $w_{i j k}^{\prime}$ denote the original and demodulated weights.
The subscripts $i,j,k$ enumerate the input channels, output channels, and spatial footpoints of the convolution weights.
$\epsilon$ is set to $1\times 10^{-8}$ to avoid numerical issues.

With the co-modulation design, our network is flexible to perform various tasks by changing the two condition coefficients $\delta$ and $c_1$.
For example, it performs CMS alone when $\delta$ is set to 0.
In this case, if the adjacent slices are unavailable, we perform one-to-one translation where only the corresponding slice is given as input.
When $\delta$ changes with the desired target position and $c_1$ is set to $c_0$, we perform SR by traversing all positions and producing 3D volumes by stacking the generated slices along the axis perpendicular to the slices.
We can also perform CMSR by setting $\delta$ and $c_1$ to the desired values.

\subsection{Alias-free Generator}
\label{sec:alias-free}
One of the major challenges in developing the generator is the aliasing artifacts in the 3D reconstructed results.
We attribute the problem to the fact that fine-grained details are unnaturally dependent on the coordinates, and therefore fixed in specific positions when $\delta$ changes. 
This phenomenon was believed to originate from aliasing caused by carelessly designed operators (\emph{i.e.}, convolution, resampling, and nonlinearity) in CNN \cite{stylegan3}.
To solve the problem, we consider the feature map $Z$ in the generator as regularly sampled signals, which can represent the corresponding continuous signal $z$ with limited frequencies under the Shannon-Nyquist signal processing framework \cite{Nyquist}.
Therefore, the discrete operation $\mathbf{F}$ on $Z$ has its continuous counterpart $\mathbf{f}(z)$:
\begin{equation}
\quad \mathbf{F}(Z)=\mathrm{\Sh}_{r_1} \odot \mathbf{f}\left(\phi_{r_0} * Z\right),
\label{alias-free}\end{equation}
where $\mathrm{\Sh}_{r_1}$ is the two-dimensional Dirac comb function with sampling rate $r_1$, and $\phi_{r_0}$ is the sinc-based interpolation filter with a band limit of $r_0/2$ so that $z$ can be represented as $z=\phi_{r_0} * Z$.
The operation $\mathbf{F}$ is alias-free when any frequencies higher than $r_1/2$ in the output signal, also known as alias frequencies, are efficiently suppressed. Following the criteria, we design an alias-free generator, which consists of the antialiasing encoder and synthesizer detailed as follows.

\subsubsection{Antialiasing Encoder}
Considering that any details with alias frequencies need to be suppressed, we refer from Alias-Free GAN \cite{stylegan3} to design the novel resampling and nonlinearity operators:
The encoder consists of 14 layers, each of which is further composed of a convolution, a nonlinear operation, and a filtered downsampling.
The convolution preserves the original form as it does not introduce new frequencies.
In low-resolution layers, the resampling filter is designed as non-critical sinc one whose cutoff frequency varies with the resolution of feature maps. 
Specifically, the cutoff frequency geometrically decreases from $f_c=r_N/2$ in the first non-critically sampled layer (\emph{i.e.,} the third layer) to $f_c=2$ in the last layer, where $r_N$ is the image resolution.
The minimum acceptable stopband frequency starts at $f_t=2^{0.3} \cdot r_N/2$ and geometrically decreases to $f_t=2^{2.1}$, whose value determines the resolution $r={\rm min}({\rm ceil}(2 \cdot f_t), r_N)$ and the transition band half-width $f_h={\rm max}(r/2, f_t) - f_c$ in each layer.
The target feature map of each layer is padded and surrounded by a 10-pixel margin, and the final feature map is resampled to 4$\times$4 before formulating the image-conditioned representation. We also modulate the convolution weights of the encoder with attribute embeddings $\mathbf{v}$ transformed by another attribute mapping network $\mathcal{N}$, so that the encoder focuses on the specific task and provides more relevant conditions.

\subsubsection{Antialiasing Synthesizer}
Following the previous work \cite{comodgan}, skip connections are implemented between $\mathcal{E}$ and $\mathcal{D}$ to preserve the anatomical structure of input images.
Therefore, the decoder is also designed to be alias-free so that skip connections would not introduce extra content with undesired frequency.
The design of the synthesizer strictly follows those of the encoder.
Each layer of the synthesizer consists of a modulated convolution, the upsampling operation, and an activation function.
Different from the encoder which directly performs the activation, we wrap the nonlinear operation (\emph{i.e.}, Leaky ReLU) between upsampling and downsampling to ensure that any high-frequency content introduced by the operation is suppressed.

\subsection{Projection Discriminator}
\label{sec:projection}
One of the most noteworthy problems when tailoring 2D-based synthesizers to 3D MR images is the inconsistency between generated slices.
In orthogonal views, the generated images can be discontinuous if such consistency is not well ensured.
However, the vanilla 2D synthesizer is guided by either the pixel-wise losses or a simple 2D discriminator, which is not effective enough to empower the synthesizer with cross-slice consistency.
To address the issue, we propose to tailor the projection discriminator \cite{cgan} to capture the consistency using class-conditioned and position-embedded representation.
As presented in \autoref{overall}, the input images $\mathbf{x}_{in}$ are concatenated with the generated slice $\hat{\mathbf{y}}^{k}$ or the real slice $\mathbf{y}^{k}$ before feeding to the discriminator, which consists of 7 residual blocks and a projection block.
We refer from StyleGAN2 \cite{StyleGAN2} to design the residual blocks. The projection block is composed of modified minibatch discrimination \cite{ProgressiveGAN}, fully connected layers, and the projection strategy that projects the embedded attribute $\mathbf{v}$.
Then the discriminator differentiates whether the image is real or generated, as well as the inter-slice consistency using the neighboring slices and the embedded attribute.

\subsection{SAM-Guided Synthesis}
\label{sec:sam}
Segmentation-guided synthesis is a popular method to impose the structure constraint and help the generative models focus on regions of interest  (ROIs) \cite{SegGuided2, SegGuided5, SegGuided4}.
Recently, the Segment Anything Model (SAM) \cite{SAM} has emerged as the first foundation model for promptable image segmentation.
Built upon the vision transformer, SAM is trained on a large dataset including more than 1 billion masks from 11 million images, and exhibits impressive zero-shot segmentation performance.
Several studies have reported that SAM helps segment anatomical structures in MRI with the appropriate prompt \cite{SAM_mri1, SAM_mri2}, which can be utilized for the image synthesis task by providing its segmentation results as strcutral priors for refining the reconstructed MR images.
However, SAM cannot be used directly for segmentation-guided synthesis, as SAM-guided segmentation results are often undesired \cite{SAM_not_perfect} and sensitive to hyperparameters for post-processing. Therefore, we instead use the ViT-based encoder $E$ of SAM to extract structure-related features for target slices. 
Such structural information from SAM is valuable as prior knowledge for reconstructing MR images, especially for those with ROIs such as clots or tumors.
Specifically, we use these features to guide the generator with the SAM loss $L_{SAM}$ as follows:
\begin{equation}
L_{SAM}=\mathbb{E}_{\hat{\mathbf{y}}^{k}, \mathbf{y}^{k}} \| E(R(\hat{\mathbf{y}}^{k})) - E(R(\mathbf{y}^{k})) \|_1 ,
\label{loss_sam}\end{equation}
where $E$ is the encoder of SAM, and $R$ is the rearrangement operator that reorganizes the tensor to match the shape required by SAM. The equation is written following the $\mathrm{einops}$ notation \cite{einops} as:
\begin{equation}
R(\mathbf{x}) = \mathrm{rearrange}(\mathbf{x}, \mathrm{b} \enspace \mathrm{c} \enspace \mathrm{h} \enspace \mathrm{w} \rightarrow \mathrm{1} \enspace \mathrm{c} \enspace \mathrm{h\sqrt{b}} \enspace \mathrm{w\sqrt{b}})
\label{rearrange}\end{equation}

\subsection{Training Details}
\label{sec:implement}
The overall loss for training \textit{Uni-COAL} is composed of an adversarial loss, a pixel-wise $L_1$ loss, and a SAM loss:
\begin{equation}
L_G=L_{adv} + \lambda_1 L_{pix} + \lambda_2 L_{SAM},
\label{loss_g}\end{equation}
where $\lambda_1$ and $\lambda_2$ are the hyper-parameters to balance the losses.
The adversarial loss $L_{adv}$ is defined as non-saturation loss with $R_1$ regularization, whose objective functions are:
\begin{equation}
\begin{aligned}
\min _{\mathcal{D}is} V(\mathcal{D}is) &= - \mathbb{E}_{\mathbf{x}_{in}, \mathbf{y}^{k}}\left[\log \mathcal{D}is(\mathbf{x}_{in}, \mathbf{y}^{k})\right] \\
&- \mathbb{E}_{\mathbf{x}_{in}, \hat{\mathbf{y}}^{k}} \left[\log(1-\mathcal{D}is(\mathbf{x}_{in}, \hat{\mathbf{y}}^{k}))\right] \\
&+\frac{{\gamma}_{g}}{2} \mathbb{E}_{\mathbf{x}_{in}, \mathbf{y}^{k}} \left[\left\|\nabla \mathcal{D}is(\mathbf{x}_{in}, \mathbf{y}^{k})\right\|^{2}\right] \\
\max _{\mathcal{G}} V(\mathcal{G}) &= \mathbb{E}_{\mathbf{x}_{in}, \hat{\mathbf{y}}^{k}} \left[\log \mathcal{D}is(\mathbf{x}_{in}, \hat{\mathbf{y}}^{k})\right].
\end{aligned}
\label{nonsaturation_loss}\end{equation}

Note that to stabilize training and facilitate convergence, we refer from \cite{stylegan3} to smooth the target and reconstructed images using a Gaussian filter whose $\sigma$ is initialized with 2 pixels and linearly decayed to 0 for the first 100$k$ images.
The exponential moving average of the generator is used for inference.

\setcounter{footnote}{0}
\section{Experiments}
\subsection{Datasets}
To comprehensively evaluate the performance of \textit{Uni-COAL} and validate its robustness under different MR reconstruction scenarios, here we conduct a series of experiments on four brain MRI datasets, which are the BRATS \cite{brats}, in-house CSDH, ADNI \cite{adni}, and IXI \cite{ixi} datasets detailed as follows:

\textbf{BRATS} is a public MRI dataset scanned from low- and high-grade glioma patients.
The MR images are obtained with diverse clinical protocols and scanners across multiple institutions.
As described in its original work, multi-modality images (\emph{i.e.}, T1, T2-weighted, post-contrast T1 weighted (T1Gd), and FLAIR) have undergone co-registration to the same anatomical template, with interpolation to a 1 mm $\times $ 1 mm $\times $ 1 mm spatial resolution, and skull stripping.
We randomly divide 335 scans in BraTS'19 challenge into 235, 50, 50 for training, validation, and testing.
The 80 middle axial slices are extracted and further padded to 256$\times$256.
Annotations for tumor regions are also used to evaluate the synthesis quality in the segmentation tasks.

\textbf{CSDH} is an in-house dataset comprising 100 patients diagnosed with chronic subdural hematoma (CSDH). 
It includes T1-weighted images scanned with repetition time (TR) = 1700 ms, echo time (TE) = 2.45 ms, and spatial resolution = 0.75 mm $\times $ 0.75 mm $\times $ 1 mm, as well as T2-FLAIR images scanned with TR = 8000 ms, TE = 83 ms, and spatial resolution = 0.75 mm $\times $ 0.75 mm $\times $ 8 mm.
Pixel-wise annotations of liquefied blood clots made by an experienced radiologist are also used for segmentation accuracy evaluation. Note that we divide the ADNI dataset into 70, 10, and 20 subjects for training, validating, and testing, respectively.

\textbf{ADNI} is composed of 50 patients diagnosed with Alzheimer's disease and 50 elderly controls following \cite{synthsr}. 
Each subject has one near-isotropic T1 MP-RAGE scan with thickness of 1 mm and one axial FLAIR scan with thickness of 5 mm. We adopt the same settings as CSDH dataset for dividing the ADNI dataset for training, validating, and testing.

\textbf{IXI} is a collection of multi-modality near-isotropic MRI images obtained from three hospitals. After rigid registration using ANTs \cite{ANTs}, we select 309 healthy subjects scanned in the Guy’s Hospital with spatial resolution of 0.9375 mm $\times $ 0.9375 mm $\times $ 1.25 mm for the experiment, which are further divided into 285 subjects for training, 12 subjects for validation, and 12 subjects for testing.
For each subject, T1, T2, and PD-weighted (PD) MR images are preserved for analysis.

Note that we conduct different experiments on these datasets according to their included modalities, slice thickness of MR images and annotation information. Specifically, we use the BRATS dataset to evaluate the CMS performance, and perform SR and CMSR on CSDH and ADNI as they contain multi-modality images with both isotropic and anisotropic volumes. 
We also design an arbitrary-modality and arbitrary-scale CMSR experiment on IXI. As CSDH and BRATS datasets have lesion annotations, we further conduct experiments by applying a pretrained segmentation model to the synthesized images, and investigate whether they have accurately preserved the anatomical structure of the lesion regions. In this way, we use the Peak Signal-to-Noise Ratio (PSNR) and Structural Similarity (SSIM) to quantify the quality of generated MR images, and the Dice Similarity Coefficient (DSC) to measure the segmentation performance. 

For SR-related tasks, we follow the previous work \cite{DeepResolve} to simulate the acquisition of high-thickness slices on the CSDH, ADNI, and IXI datasets. Specifically, the original images are first filtered to simulate the magnetization contributions from surrounding slices, and then downsampled to the target resolution.

\subsection{Implementation Details}
\textit{Uni-COAL} is trained with batch size 16 on an NVIDIA GeForce RTX 3090 GPU for 100 epochs. It takes about 36 GPU hours for training and 5 seconds per 3D image for inference.
The learning rates are initialized as 0.0025/0.0020 for the generator and the discriminator respectively during the first 50 epochs, and then linearly decrease to 0. Also note that we use the translation equivariant configuration of the Alias-Free GAN, and discard the rotational equivariance to avoid generating overly-symmetric images.

\begin{table*}[ht]
\centering
\caption{\label{tab:cm_table}
Quantitative results for cross-modality synthesis on the BRATS dataset.}
\resizebox{\textwidth}{!}{%
\begin{tabular}{lccccccccccc}
\thickhline
\multicolumn{1}{c}{\multirow{2}{*}{}} & \multicolumn{3}{c}{T1-T2} &  & \multicolumn{3}{c}{T2-FLAIR} &  & \multicolumn{3}{c}{FLAIR-T1} \\ \cline{2-4} \cline{6-8} \cline{10-12} 
\multicolumn{1}{c}{} & PSNR & SSIM & DSC &  & PSNR & SSIM & DSC &  & PSNR & SSIM & DSC \\ \hline
\multicolumn{12}{l}{\textbf{2D-based Cross-modality Synthesis}} \\ 
pix2pix \cite{pix2pix} & 25.10$\pm$1.54 & 0.912$\pm$0.021 & 0.632$\pm$0.223 &  & 22.89$\pm$1.73 & 0.867$\pm$0.017 & 0.785$\pm$0.134 &  & 23.60$\pm$1.42 & 0.888$\pm$0.020 & 0.675$\pm$0.162 \\
AutoGAN \cite{AutoGAN} & 25.57$\pm$1.87 & 0.910$\pm$0.020 & 0.695$\pm$0.192 &  & 24.10$\pm$2.00 & 0.870$\pm$0.018 & 0.777$\pm$0.127 &  & 24.16$\pm$1.40 & 0.881$\pm$0.020 & 0.721$\pm$0.143 \\
ResViT \cite{ResViT} & 26.16$\pm$1.89 & 0.923$\pm$0.022 & 0.737$\pm$0.157 &  & 24.94$\pm$1.85 & 0.891$\pm$0.019 & 0.833$\pm$0.092 &  & 24.74$\pm$1.49 & 0.906$\pm$0.021 & 0.747$\pm$0.146 \\
\textit{Uni-COAL}$_{\rm one}$ & 26.18$\pm$2.07 & 0.928$\pm$0.022 & 0.731$\pm$0.192 &  & 24.83$\pm$1.80 & 0.896$\pm$0.019 & 0.834$\pm$0.095 &  & 25.00$\pm$1.68 & 0.913$\pm$0.021 & 0.752$\pm$0.129 \\ \hline
\multicolumn{12}{l}{\textbf{2.5D or 3D-based Cross-modality Synthesis}} \\ 
pGAN \cite{pGAN} & 26.28$\pm$1.88 & 0.927$\pm$0.022 & 0.758$\pm$0.144 &  & 24.94$\pm$1.93 & 0.893$\pm$0.018 & 0.841$\pm$0.092 &  & 25.00$\pm$1.64 & 0.911$\pm$0.021 & \textbf{0.754$\pm$0.151} \\
SC-GAN \cite{SCGAN} & 25.10$\pm$1.78 & 0.917$\pm$0.021 & 0.669$\pm$0.216 &  & 23.20$\pm$2.63 & 0.858$\pm$0.024 & 0.788$\pm$0.157 &  & 23.38$\pm$1.66 & 0.887$\pm$0.018 & 0.656$\pm$0.170 \\
ProvoGAN \cite{ProvoGAN} & 26.12$\pm$1.54 & 0.922$\pm$0.024 & 0.721$\pm$0.169 &  & 24.35$\pm$1.81 & 0.884$\pm$0.019 & 0.810$\pm$0.142 &  & 24.57$\pm$1.48 & 0.904$\pm$0.023 & 0.739$\pm$0.152 \\
\textit{Uni-COAL}$_{\rm multi}$ & \textbf{26.47$\pm$2.22} & \textbf{0.932$\pm$0.023} & \textbf{0.761$\pm$0.194} &  & \textbf{25.04$\pm$1.92} & \textbf{0.900$\pm$0.018} & \textbf{0.848$\pm$0.071} &  & \textbf{25.24$\pm$1.79} & \textbf{0.918$\pm$0.021} & 0.749$\pm$0.123 \\
\thickhline
\end{tabular}%
}
\end{table*}

\begin{figure*}[t]
\includegraphics[width=\textwidth]{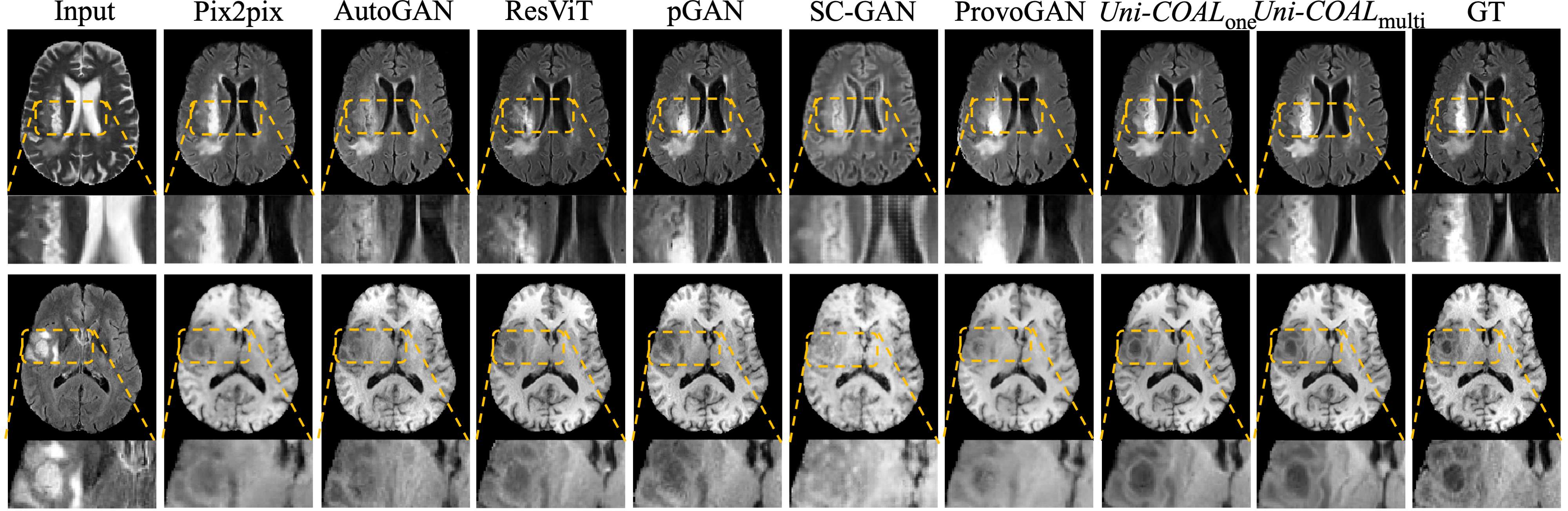}
\caption{Qualitative results for cross-modality synthesis on the BRATS dataset.} \label{results_cm}
\end{figure*}

\subsection{Comparative Experiments}
\subsubsection{Cross-Modality Synthesis}
In this section, we report three subtasks, namely CMS from T1 to T2, T2 to FLAIR, and FLAIR to T1-weighted MR images on the BRATS dataset.
As shown in \autoref{tab:cm_table}, both 2D-based CMS and 2.5D/3D-based CMS methods are performed for a comprehensive comparison.
In terms of 2D-based CMS, \textit{Uni-COAL}$_{\rm one}$, which takes the one corresponding slice as the input, achieves significantly superior results for all tasks compared to pix2pix and AutoGAN (p$<$0.05) and matches the performance with ResViT which is the SOTA method for 2D CMS. The improved generation quality can be attributed to the alias-free design of \textit{Uni-COAL}, which can accurately restore high-frequency details (\emph{e.g.,} textures of the enhancing tumor and the necrotic tumor core in \autoref{results_cm}). Moreover, the generation quality is further improved when adjacent slices are used as extra guidance, as shown in 2.5D/3D-based CMS.
In such case, \textit{Uni-COAL}$_{\rm multi}$ achieves the best results for all tasks, indicating its superiority in capturing context information.

\subsubsection{Super-Resolution}
The SR experiment is performed using the CSDH dataset with downsampling factor (DSF) of 8 and the ADNI dataset with DSF of 5.
Note that it is generally counter-intuitive to generate 2D images slice-by-slice instead of reconstructing the entire 3D volume directly.
However, results in \autoref{tab:sr_table} suggest that \textit{Uni-COAL}$_{\rm sr}$ yields higher SR quality than other 3D-based SR models due to the continuous position-embedded representation and the alias-free design for detail-restoration.
Although the improvement is not significant compared to the SOTA methods, the qualitative results in \autoref{results_sr} demonstrate that \textit{Uni-COAL}$_{\rm sr}$ produces coherent details. In contrast, the results obtained from alternative 3D-based methods are blurred due to imperfect pre-resampling, especially when the DSF is high.

We also investigate whether the synthesized high-resolution images can be used for downstream tasks. As reported in \autoref{tab:sr_table}, when using a pre-trained segmentation model to delineate the liquefied blood clots in the reconstructed images, \textit{Uni-COAL}$_{\rm sr}$ produces the most reliable results, which also indicates the superiority of our method for clinical applications.

\subsubsection{Cross-Modality Synthesis and Super-Resolution}
\begin{figure*}[t]
\includegraphics[width=\textwidth]{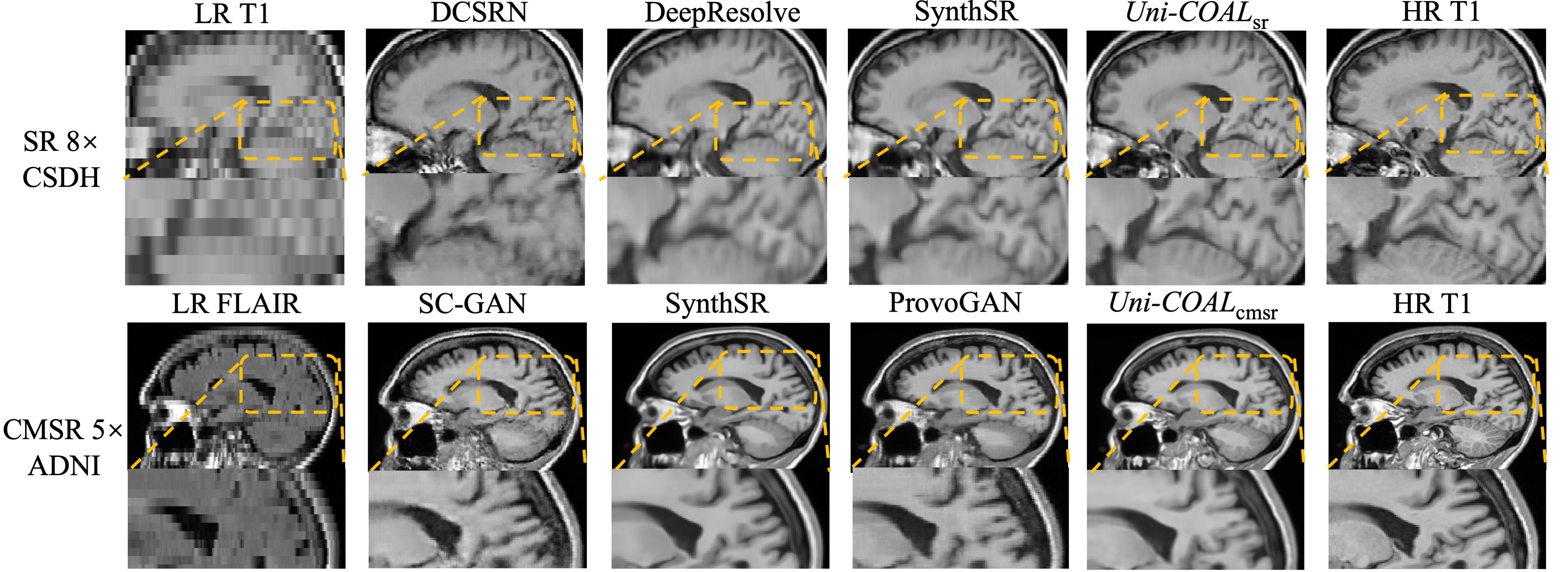}
\caption{Qualitative results for SR and CMSR on the CSDH and ADNI datasets.} \label{results_sr}
\end{figure*}

We evaluate on the CSDH and ADNI dataset by synthesizing high-resolution T1 images given low-resolution FLAIR images.
As reported in \autoref{tab:cmsr_table}, \textit{Uni-COAL} significantly outperformed other 3D-based methods (p$<$0.05).
This can be attributed to the superiority of our method for capturing and restoring modality-invariant details more accurately with the co-modulation design.
The qualitative results in \autoref{results_sr} further demonstrate that the alternative 3D-based CMSR methods exhibit blurriness (SC-GAN and SynthSR) or artifacts (SC-GAN and ProvoGAN), while \textit{Uni-COAL}$_{\rm cmsr}$ produces high-quality results with notable fidelity.
Similar to the SR experiment, the MR images reconstructed by \textit{Uni-COAL}$_{\rm cmsr}$ exhibit more accurate preservation of anatomical structures, which are also confirmed by the highest DSC.

\begin{table}[t]
\centering
\caption{\label{tab:sr_table}
Quantitative results for super-resolution on the CSDH and ADNI datasets. }
\resizebox{\columnwidth}{!}{%
\begin{tabular}{lcccccc}
\thickhline
\multicolumn{1}{c}{\multirow{2}{*}{}} & \multicolumn{3}{c}{CSDH (8$\times$)} &  & \multicolumn{2}{c}{ADNI (5$\times$)} \\ \cline{2-4} \cline{6-7} 
\multicolumn{1}{c}{} & PSNR & SSIM & DSC &  & PSNR & SSIM \\ \hline
DCSRN\cite{DCSRN} & 24.15$\pm$0.67 & 0.815$\pm$0.012 & 0.818$\pm$0.118 &  & 27.25$\pm$0.75 & 0.911$\pm$0.012 \\
DeepResolve\cite{DeepResolve} & 27.39$\pm$0.61 & 0.906$\pm$0.009 & 0.876$\pm$0.078 &  & 28.82$\pm$0.69 & 0.945$\pm$0.009 \\
SynthSR\cite{synthsr} & 28.00$\pm$0.72 & 0.917$\pm$0.013 & 0.883$\pm$0.063 &  & 29.11$\pm$0.69 & 0.936$\pm$0.009 \\
\textit{Uni-COAL}$_{\rm sr}$ & \textbf{28.18$\pm$1.02} & \textbf{0.924$\pm$0.016} & \textbf{0.896$\pm$0.056} &  & \textbf{29.14$\pm$0.69} & \textbf{0.952$\pm$0.011} \\
\thickhline
\end{tabular}%
}
\end{table}

\subsubsection{Arbitrary Cross-Modality Synthesis and Super-Resolution}
\begin{figure}[t]
\includegraphics[width=\columnwidth]{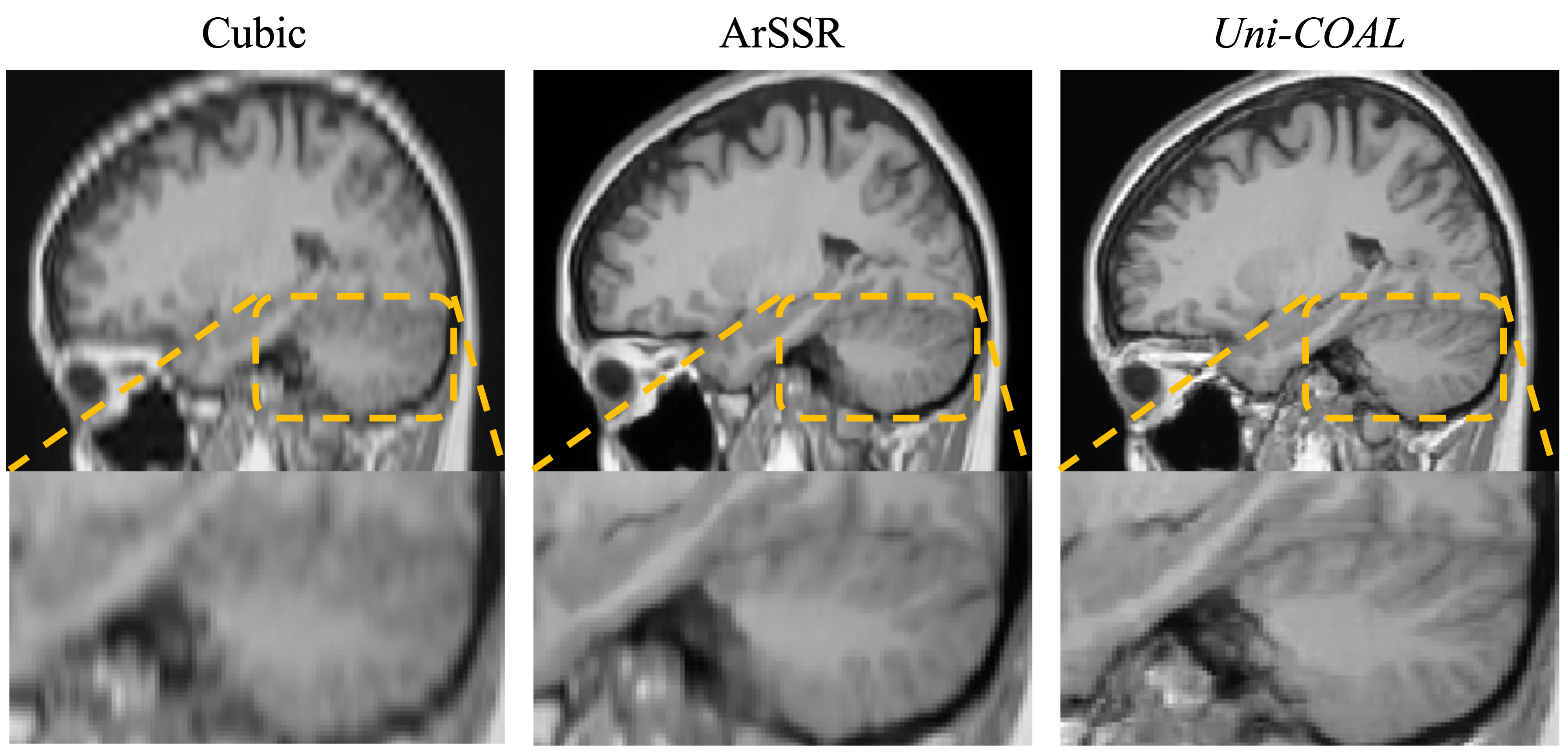}
\caption{Qualitative results for arbitrary-scale super-resolution of an low-resolution T1 MR image with spatial resolution of 0.9375 mm $\times $ 0.9375 mm $\times $ 5.0 mm and upsampling scale of 5.33$\times$.} \label{non_integer}
\end{figure}

We also conduct experiments to further evaluate the arbitrary properties of our method, which is prone to synthesizing MR images with various modality and slice thickness settings.
To this end, we experiment with the IXI dataset which contains near-isotropic images with three different modalities.
In the training stage, we randomly select the input modalities and downsample them with DSF of 2, 4, or 6.
The output MR images are also randomly selected among the three modalities with near-isotropic volumes, which can be determined with positional embeddings and class embeddings.
Results in \autoref{tab:arbitrary} indicate that although all tasks are performed with a single trained model, \textit{Uni-COAL} outperforms other methods in most cases, including SC-GAN and ProvoGAN that train a specific model for each task.
We also demonstrate the superiority over other multitask models such as ResViT, ArSSR, and SynthSR.
In addition, we qualitatively evaluate whether our network is able to perform SR or CMSR with non-integer target thickness.
To this end, we directly use the trained model to generate HR images with thickness of 0.9375 mm (\emph{i.e.}, 5.33$\times$ SR/CMSR). 
As illustrated in \autoref{intro} and \autoref{non_integer}, the \textit{Uni-COAL} model still generates reasonable results, demonstrating the effectiveness of reducing the thickness to flexible values, which were not explicitly included during the training process.

\subsection{Ablation Study}
In this section, we evaluate the impact of each component of our framework by performing CMSR on the CSDH dataset.
First, we evaluate our alias-free design, including the anti-aliasing encoder and the decoder.
Then, we assess whether our projection discriminator strategy helps guarantee the inter-slice consistency and its performance gain.
Furthermore, we measure the effectiveness of the SAM-guided strategy in \textit{Uni-COAL} framework.
The experimental results are presented in \autoref{tab:ablation} and \autoref{ablation}.

\begin{table}[t]
\centering
\caption{\label{tab:cmsr_table}
Quantitative results for cross-modality synthesis and super-resolution on the CSDH and ADNI datasets. }
\resizebox{\columnwidth}{!}{%
\begin{tabular}{lcccccc}
\thickhline
\multicolumn{1}{c}{\multirow{2}{*}{}} & \multicolumn{3}{c}{CSDH (8$\times$)} &  & \multicolumn{2}{c}{ADNI (5$\times$)} \\ \cline{2-4} \cline{6-7} 
\multicolumn{1}{c}{} & PSNR & SSIM & DSC &  & PSNR & SSIM \\ \hline
SC-GAN\cite{SCGAN} & 21.73$\pm$0.92 & 0.751$\pm$0.035 & 0.789$\pm$0.167 &  & 22.00$\pm$1.00 & 0.762$\pm$0.027 \\
SynthSR\cite{synthsr} & 22.65$\pm$0.76 & 0.815$\pm$0.021 & 0.819$\pm$0.091 &  & 23.84$\pm$1.48 & 0.822$\pm$0.028 \\
ProvoGAN\cite{ProvoGAN} & 23.24$\pm$1.00 & 0.831$\pm$0.034 & 0.831$\pm$0.114 &  & 24.10$\pm$0.98 & 0.849$\pm$0.021 \\
\textit{Uni-COAL}$_{\rm cmsr}$ & \textbf{23.53$\pm$1.52} & \textbf{0.850$\pm$0.036} & \textbf{0.859$\pm$0.071} &  & \textbf{24.19$\pm$1.37} & \textbf{0.866$\pm$0.024} \\
\thickhline
\end{tabular}%
}
\end{table}

\subsubsection{Alias-free Design}
We replace the encoder and decoder in \textit{Uni-COAL} with the vanilla ones \cite{comodgan} as baseline for comparison.
Results in \autoref{tab:ablation} indicate that while redesigning the decoder alone resulted in the improvement in SSIM, it is the simultaneous redesign of both encoder and decoder that leads to the increase of PSNR and DSC, which highlights the significance of redesigning both components to achieve optimal results.


\begin{table*}[t]
\centering
\caption{\label{tab:arbitrary}
Quantitative results for arbitrary cross-modality synthesis and super-resolution. }
\resizebox{\textwidth}{!}{%
\begin{tabular}{lcccccccccccc}
\thickhline
\multirow{2}{*}{Scale} & \multicolumn{2}{c}{ResViT} & \multicolumn{2}{c}{ArSSR} & \multicolumn{2}{c}{SC-GAN} & \multicolumn{2}{c}{ProvoGAN} & \multicolumn{2}{c}{SynthSR} & \multicolumn{2}{c}{\textit{Uni-COAL}} \\
 & PSNR & SSIM & PSNR & SSIM & PSNR & SSIM & PSNR & SSIM & PSNR & SSIM & PSNR & SSIM \\ \hline
\multicolumn{13}{l}{\textbf{T1 SR}} \\
2$\times$ & - & - & 33.47$\pm$0.59 & 0.980$\pm$0.002 & 33.86$\pm$0.60 & 0.985$\pm$0.001 & 34.41$\pm$0.69 & 0.987$\pm$0.001 & 31.22$\pm$0.55 & 0.967$\pm$0.003 & \textbf{34.91$\pm$0.68} & \textbf{0.989$\pm$0.001} \\
4$\times$ & - & - & 26.41$\pm$0.64 & 0.948$\pm$0.007 & 27.28$\pm$0.51 &0.945$\pm$0.006 & 29.38$\pm$0.65 & 0.962$\pm$0.004 & 28.10$\pm$0.57 & 0.944$\pm$0.007 & \textbf{30.34$\pm$0.63} & \textbf{0.968$\pm$0.003} \\
6$\times$ & - & - & 26.36$\pm$0.57 & 0.900$\pm$0.010 & 26.02$\pm$0.43 &0.897$\pm$0.009 & 27.02$\pm$0.52 & 0.912$\pm$0.009 & 26.40$\pm$0.62 & 0.895$\pm$0.011 & \textbf{28.18$\pm$0.58} & \textbf{0.947$\pm$0.005} \\ \hline
\multicolumn{13}{l}{\textbf{T1-T2 CMS/CMSR}} \\
1$\times$ & 26.37$\pm$1.09 & 0.914$\pm$0.017 & - & - & 23.77$\pm$0.77 & 0.884$\pm$0.016 & 25.55$\pm$0.95 & 0.903$\pm$0.016 & 26.01$\pm$1.08 & 0.908$\pm$0.016 & \textbf{26.43$\pm$1.06} & \textbf{0.918$\pm$0.013} \\
2$\times$ & - & - & - & - & 22.50$\pm$0.60 &0.876$\pm$0.017 & 25.02$\pm$0.77 & 0.899$\pm$0.015 & 25.92$\pm$1.03 & 0.908$\pm$0.015 & \textbf{26.35$\pm$0.99} & \textbf{0.917$\pm$0.013} \\
4$\times$ & - & - & - & - & 22.27$\pm$0.61 & 0.843$\pm$0.019 & 24.93$\pm$0.77 & 0.887$\pm$0.015 & 25.20$\pm$0.90 & 0.892$\pm$0.015 & \textbf{25.50$\pm$0.84} & \textbf{0.903$\pm$0.016} \\
6$\times$ & - & - & - & - & 21.92$\pm$0.72 & 0.841$\pm$0.019 & 24.34$\pm$0.71 & 0.874$\pm$0.015 & 24.47$\pm$0.78 & 0.875$\pm$0.015 & \textbf{24.50$\pm$0.71} & \textbf{0.884$\pm$0.015} \\ \hline
\multicolumn{13}{l}{\textbf{T2 SR}} \\
2$\times$ & - & - & 33.40$\pm$0.79 & 0.978$\pm$0.003 & 33.88$\pm$0.81 & 0.976$\pm$0.004 & 34.04$\pm$0.88 & 0.979$\pm$0.004 & 34.74$\pm$1.01 & 0.982$\pm$0.003 & \textbf{35.00$\pm$0.97} & \textbf{0.983$\pm$0.003} \\
4$\times$ & - & - & 26.17$\pm$0.83 & 0.929$\pm$0.008 & 27.32$\pm$0.54 & 0.931$\pm$0.007 & 28.94$\pm$0.59 & 0.946$\pm$0.005 & 29.50$\pm$0.66 & 0.949$\pm$0.006 & \textbf{29.76$\pm$0.68} & \textbf{0.955$\pm$0.006} \\
6$\times$ & - & - & 25.38$\pm$0.58 & 0.905$\pm$0.011 & 25.88$\pm$0.57 & 0.908$\pm$0.010 & 26.87$\pm$0.54 & 0.916$\pm$0.009 & \textbf{27.31$\pm$0.62} & 0.923$\pm$0.008 & 27.13$\pm$0.63 & \textbf{0.930$\pm$0.008} \\ \hline
\multicolumn{13}{l}{\textbf{T2-PD CMS/CMSR}} \\
1$\times$ & 31.54$\pm$0.48 & 0.950$\pm$0.006 & - & - & 29.06$\pm$0.50 & 0.947$\pm$0.007 & 30.49$\pm$0.58 & 0.957$\pm$0.006 & 28.80$\pm$0.75 & 0.940$\pm$0.013 & \textbf{32.02$\pm$0.57} & \textbf{0.968$\pm$0.005} \\
2$\times$ & - & - & - & - & 28.26$\pm$0.61 & 0.925$\pm$0.007 & 29.99$\pm$0.46 & 0.949$\pm$0.006 & 27.94$\pm$0.71 & 0.932$\pm$0.013 & \textbf{30.70$\pm$0.70} & \textbf{0.961$\pm$0.006} \\
4$\times$ & - & - & - & - & 26.50$\pm$0.56 & 0.918$\pm$0.010 & 28.06$\pm$0.74 & 0.929$\pm$0.008 & 26.15$\pm$0.68 & 0.907$\pm$0.014 & \textbf{28.26$\pm$0.63} & \textbf{0.942$\pm$0.008} \\
6$\times$ & - & - & - & - & 26.12$\pm$0.57 & 0.899$\pm$0.012 & 26.10$\pm$0.72 & 0.901$\pm$0.009 & 24.85$\pm$0.70 & 0.891$\pm$0.012 & \textbf{26.44$\pm$0.66} & \textbf{0.920$\pm$0.011} \\ \hline
\multicolumn{13}{l}{\textbf{PD SR}} \\
2$\times$ & - & - & 33.90$\pm$0.69 & 0.981$\pm$0.003 & 34.72$\pm$0.81 & 0.982$\pm$0.003 & 35.22$\pm$0.48 & 0.985$\pm$0.003 & 32.67$\pm$0.76 & 0.970$\pm$0.003 & \textbf{35.60$\pm$0.96} & \textbf{0.986$\pm$0.002} \\
4$\times$ & - & - & 25.88$\pm$0.73 & 0.928$\pm$0.006 & 28.22$\pm$0.57 & 0.937$\pm$0.006 & 29.60$\pm$0.66 & 0.950$\pm$0.005 & 29.05$\pm$0.66 & 0.940$\pm$0.005 & \textbf{30.69$\pm$0.68} & \textbf{0.961$\pm$0.004} \\
6$\times$ & - & - & 25.56$\pm$0.68 & 0.908$\pm$0.009 & 26.58$\pm$0.59 & 0.913$\pm$0.009 & 27.03$\pm$0.65 & 0.922$\pm$0.008 & 26.96$\pm$0.60 & 0.918$\pm$0.008 & \textbf{28.20$\pm$0.65} & \textbf{0.942$\pm$0.007} \\ \hline
\multicolumn{13}{l}{\textbf{PD-T1 CMS/CMSR}} \\
1$\times$ & 26.32$\pm$1.17 & 0.903$\pm$0.018 & - & - & 24.66$\pm$0.89 & 0.901$\pm$0.011 & 26.61$\pm$0.83 & 0.927$\pm$0.013 & 25.43$\pm$0.89 & 0.912$\pm$0.016 & \textbf{27.51$\pm$1.16} & \textbf{0.948$\pm$0.011} \\
2$\times$ & - & - & - & - & 24.88$\pm$0.79 & 0.900$\pm$0.012 & 26.04$\pm$0.68 & 0.919$\pm$0.014 & 25.26$\pm$0.88 & 0.911$\pm$0.016 & \textbf{27.39$\pm$1.02} & \textbf{0.947$\pm$0.011} \\
4$\times$ & - & - & - & - & 24.12$\pm$0.55 & 0.875$\pm$0.013 & 25.02$\pm$0.77 & 0.893$\pm$0.013 & 24.63$\pm$0.76 & 0.887$\pm$0.017 & \textbf{26.71$\pm$0.90} & \textbf{0.936$\pm$0.011} \\
6$\times$ & - & - & - & - & 23.66$\pm$0.69 & 0.889$\pm$0.015 & 24.68$\pm$0.74 & 0.894$\pm$0.015 & 24.19$\pm$0.63 & 0.873$\pm$0.018 & \textbf{25.92$\pm$0.82} & \textbf{0.921$\pm$0.012} \\
 \thickhline
\end{tabular}
}
\end{table*}

\subsubsection{Projection Discriminator}
We ablate on the projection discriminator strategy by removing the dot production operation in the projection block of the discriminator.
The qualitative results reveal that the strategy effectively maintains consistency between the generated slices, as evidenced by the alleviated artifacts in the sagittal view.
Moreover, the use of the projection discriminator leads to improvements in PSNR and SSIM metrics, while DSC is slightly decreased.

\subsubsection{SAM-guided Strategy}
\begin{figure}[t]
\includegraphics[width=\columnwidth]{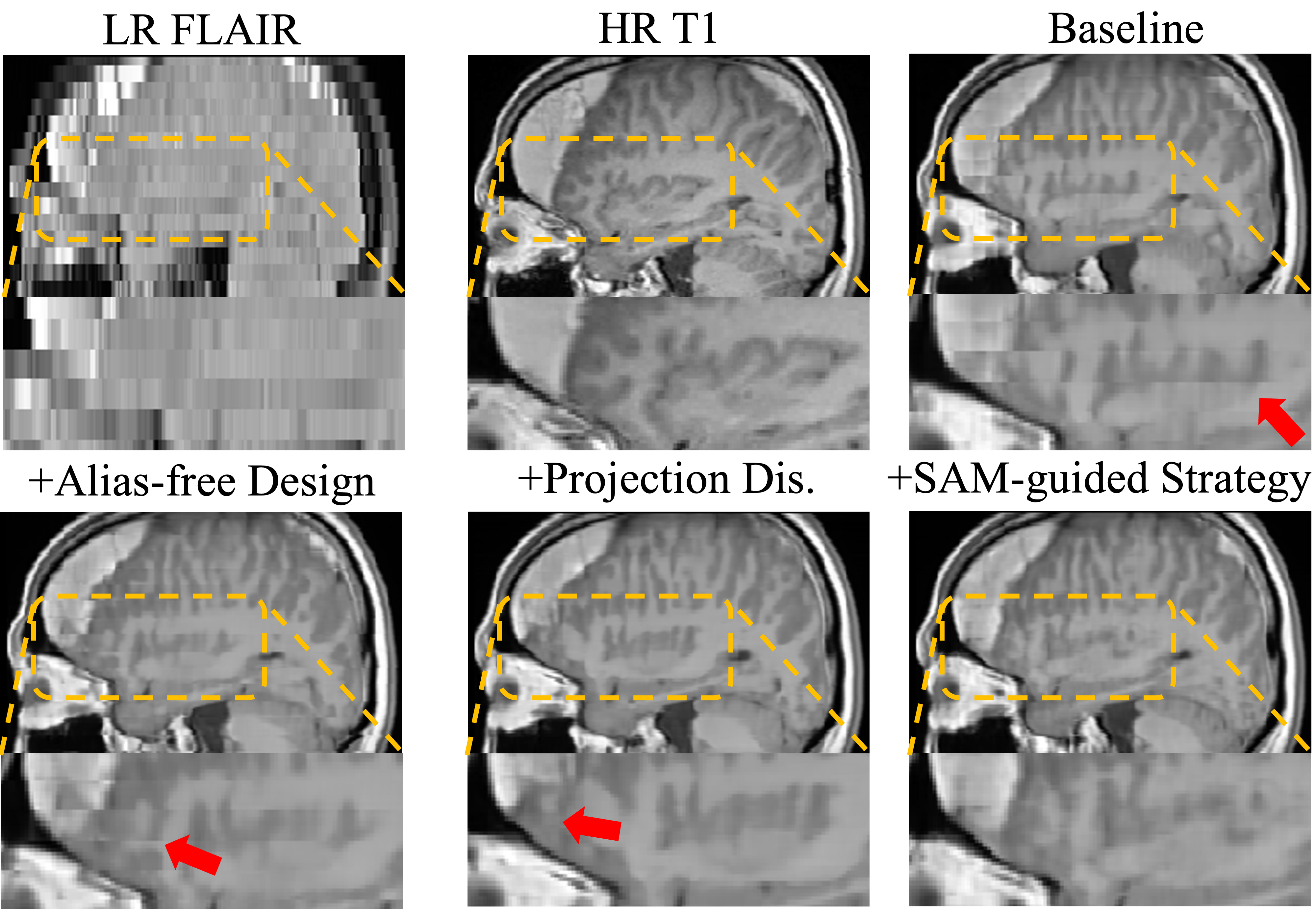}
\caption{Qualitative results for ablation study.} \label{ablation}
\end{figure}

We also explore whether the SAM model provides effective guidance for the generator. The quantitative results shown in \autoref{tab:ablation} demonstrate the improvements using the SAM-guided strategy, particularly in terms of DSC which increased from 0.844 to 0.859.
The qualitative results in \autoref{ablation} also indicate that the anatomical structures are preserved with higher quality when the extra knowledge of semantic-prior is imposed for the generator.

\begin{table}[t]
\centering
\caption{\label{tab:related_work}
The summary of the relevant work for CMS, SR, or CMSR with various input/output combinations. Spec. stands for "specific" that the model is trained for the specific input/output combination. Arb. stands for "arbitrary" that the model can deal with arbitrary input sencarios or produce arbitrary outputs. }
\resizebox{\columnwidth}{!}{%
\begin{tabular}{cccccccc}
\thickhline
\multicolumn{2}{c}{Tasks} & ResViT & ArSSR & SC-GAN & ProvoGAN & SynthSR & \textit{Uni-COAL} \\
\hline
\multirow{3}{*}{CMS} & Spec. & \checkmark &  & \checkmark & \checkmark & \checkmark & \checkmark \\
 & Arb. In & \checkmark &  &  &  & \checkmark & \checkmark \\
 & Arb. Out & \checkmark &  &  &  &  & \checkmark \\
 \hline
\multirow{3}{*}{SR} & Spec. &  & \checkmark & \checkmark & \checkmark & \checkmark & \checkmark \\
 & Arb. In &  & \checkmark &  &  & \checkmark & \checkmark \\
 & Arb. Out &  & \checkmark &  &  &  & \checkmark \\
 \hline
\multirow{3}{*}{CMSR} & Spec. &  &  & \checkmark & \checkmark & \checkmark & \checkmark \\
 & Arb. In &  &  &  &  & \checkmark & \checkmark \\
 & Arb. Out &  &  &  &  &  & \checkmark \\
 \thickhline
\end{tabular}
}
\end{table}

\begin{table}[t]
\centering
\caption{\label{tab:ablation}
Quantitative results for ablation studies. Experiments are all performed with simultaneous cross-modality synthesis and super-resolution on the CSDN dataset. }
\resizebox{\columnwidth}{!}{%
\begin{tabular}{lccc}
\thickhline
\multicolumn{1}{c}{\multirow{2}{*}{Configurations}} & \multicolumn{3}{c}{CSDH (CMSR, 8x)} \\
\multicolumn{1}{c}{} & PSNR & SSIM & DSC \\
\hline
Baseline & 22.49$\pm$1.23 & 0.823$\pm$0.037 & 0.840$\pm$0.081 \\
+Alias-free Decoder & 22.45$\pm$1.33 & 0.835$\pm$0.033 & 0.837$\pm$0.085 \\
+Alias-free Encoder & 22.59$\pm$1.27 & 0.837$\pm$0.030 & 0.846$\pm$0.078 \\
+Projection Discriminator & 23.31$\pm$1.81 & 0.843$\pm$0.035 & 0.844$\pm$0.087 \\
+SAM-guided Strategy & \textbf{23.53$\pm$1.52} & \textbf{0.850$\pm$0.036} & \textbf{0.859$\pm$0.071} \\
\thickhline
\end{tabular}
}
\end{table}

\section{Conclusion}
In this paper, we propose a unified network namely \textit{Uni-COAL} for CMS, SR, and CMSR of MR images.
It is designed to address the issues of task inconsistency between CMS and SR with a novel co-modulated design, and suppresses aliasing artifacts by an alias-free generator. 
\textit{Uni-COAL} exhibits flexibility in reconstructing MR images with non-integer upsampling scales and arbitrary modalities, which is facilitated by the specially designed attribute representation.
The incorporation of SAM-guided strategy in \textit{Uni-COAL} enhances its ability in refining anatomical structure of MR images during generation.
The experiments on four independent datasets demonstrate our state-of-the-art performance in CMS, SR, and CMSR of MR images.
In the future, we aim to expand the applicability of our method beyond brain-related tasks and explore its compatibility with alternative imaging modalities beyond MRI. We hope that our \textit{Uni-COAL} method can make a valuable contribution to the advancement of unified models in the field of medical imaging enhancement.

\bibliographystyle{IEEEtran}
\bibliography{main}

\end{document}